\documentclass[%
aip,
 jmp,%
 amsmath,amssymb,
reprint,%
]{revtex4-1}

\usepackage{graphicx}
\usepackage{dcolumn}
\usepackage{bm}

\begin{document}

\preprint{AIP/123-QED}

\title[Significance of polarization charge and isomagnetic surface]{On the significance of polarization charge and isomagnetic surface in the interaction between conducting fluid and magnetic field}

\author{Zhu-Xing Liang}
  \email{zx.liang55@gmail.com}

\author{Yi Liang}
\affiliation{18-4-102 Shuixiehuadu, Zhufengdajie, Shijiazhuang, Hebei 050035, China}

\date{\today}

\begin{abstract}

From the frozen-in field lines concept, a highly conducting fluid can move freely along, but not traverse to, magnetic field lines.
We discuss this topic and find that in the study of the frozen-in field lines concept, the effects of inductive and capacitive reactance have been omitted.
When admitted, the relationships among the motional electromotive field, the induced electric field, the eddy electric current, and the magnetic field becomes clearer and the frozen-in field line concept can be reconsidered.
We emphasize the importance of isomagnetic surfaces and polarization charges, and show analytically that whether a conducting fluid can freely traverse magnetic field lines or not depends solely on the magnetic gradient in the direction of fluid motion.
If a fluid does not change its density distribution and shape (can be regarded as a quasi-rigid body), and as long as it is moving along an isomagnetic surface, it can freely traverse magnetic field lines without any magnetic resistance no matter how strong the magnetic field is. When our analysis is applied, the origin of the magnetic field  of sunspots can be interpreted easily. In addition, we also present experimental results to support our analysis.

\end{abstract}

\maketitle

\section{Introduction}
The physical properties of magnetohydrodynamics (MHD) -- the frozen-in property of field lines and the conservation of magnetic flux\citep{Alfven42} -- are often referred to collectively as Alfv\'{e}n's Theorem. The concept of frozen-in field lines has been presented in various forms. A common interpretation is that a highly conducting fluid can move freely along, but cannot freely traverse, magnetic field lines. Because the frozen-in concept is of prime importance to plasma physics, it is widely mentioned in the literature, \citep{Ivonin98,Lundin05} and is presented in almost all textbooks of plasma physics.\citep{Boyd03,Bittencourt04,Li06,Ma12,Zheng09}
Although the applicability of this concept has long been contested,\citep{Lundin05,Akasofu78} and Alfv\'{e}n, the concept's founder, also criticized its use and stated that it is often misleading\citep{Alfven58,Alfven76}, it still plays an important role in astrophysics. For example, pulsar electrodynamics proposes that magnetosphere particles are threaded by magnetic field lines and rotate rigidly with a star.\citep{GJ69}
Here we present experimental and analytical evidence to support a conclusion that the concept of frozen-in field lines should not be used in the dynamics study of plasma.

It has been established that under certain conditions, plasma can freely cross magnetic field lines.\citep{Bellan06} Here we develop and extend the viewpoint of Bellan.

\section{\label{section2}Analysis on the interaction between conducting fluid and magnetic field lines}

First, for convenience, the term ``fluid" used in the following means an ideal conducting fluid.

Although the concept of frozen-in field lines is a ``bed-rock" concept underlying ideal MHD, it currently lacks a specific definition. The following two definitions are familiar to many readers (a discussion of multifarious definitions will be deferred to Table~\ref{tab:tableIII}):

\begin{enumerate}
\item The magnetic flux through a fluid is conserved (we designate this definition ``conservation theorem").
\item Any motion of the fluid perpendicular to the field lines carries the field lines with the fluid (we designate this definition ``frozen theorem"). This frozen theorem is often presented in another form: ``a fluid cannot freely traverse magnetic field lines."
\end{enumerate}

 A magnetic flux can be conserved in both scenarios:
\begin{enumerate}
\item All field lines within a fluid move along with the fluid.
\item The number of field lines entering a given fluid equals that leaving it at all times.
\end{enumerate}
Hence the two theorems are actually not equivalent. We accept the conservation theorem but question the frozen theorem.

There is a corollary to the frozen theorem: At low beta where the magnetic field is strong and produced by external coils, one pictures the fluid being pushed around by the magnetic field (so if the field is stationary, so is the fluid), while at high beta where the magnetic field is weak, one pictures the fluid pushing the magnetic field around. In either case, the magnetic field and the fluid move together.
Note that this corollary and the frozen theorem include two notions:

\begin{enumerate}
\item Under conditions that the external magnetic body and its magnetic field are stationary, if a fluid moves perpendicularly to the field, the magnetic field lines must bend and exert a resistive force on the fluid .
\item The magnitude of the resistive force exerted on the fluid depends on the magnetic field strength. The stronger the magnetic field, the greater the resistive force.
\end{enumerate}

It is indisputable that under general conditions when a fluid traverses a magnetic field, a resistive force will be exerted on the fluid. However, this does not mean that the traverse motion must result in the appearance of a magnetic resistance. In this paper, we seek those particular conditions under which a fluid can freely traverse magnetic field without any magnetic resistance, no matter how strong the magnetic field is.

\subsection{Conditions for fluid's freely traversing magnetic field lines}

The frozen theorem follows from the evolution equation of the magnetic field (also referred to as frozen-in field equation)
\begin{equation}
\frac{\partial{\textbf{B}}}{\partial{t}}=\nabla\times
(\textbf{u}\times\textbf{B}) , \label{e1}
\end{equation}
where \textbf{u} is the relative velocity of the fluid element moving through the magnetic field, and \textbf{u}$\times$\textbf{B} is the motional electromotive field.
Equation (1) describes the evolution (namely turbulence) of the magnetic field with fluid motion traversing the magnetic field.

From Equation (1), we realize that
\begin{enumerate}
\item For any spatial position passed by a fluid, the magnetic evolution can arise only from field-line crossings of the fluid. In its absence, there must be no magnetic evolution.
\item As long as $\nabla\times
(\textbf{u}\times\textbf{B})=0$, the magnetic evolution is absent, even if $\textbf{u}\times\textbf{B}\neq0$.
\end{enumerate}
This indicates that field-line crossing is a necessary, but not sufficient, condition for magnetic evolution. The magnetic evolution arising from the motion of the fluid always co-exists with magnetic resistance and neither phenomenon can appear alone because both arise from one and the same eddy current. Therefore, field-line crossing is also a necessary but not sufficient condition for magnetic resistance.
In other words, even if fluid crosses a magnetic field, i.e. \textbf{u}$\times$\textbf{B}$\neq0$, the magnetic evolution and magnetic resistance can be absent as long as the condition $\nabla\times$(\textbf{u}$\times$\textbf{B})=0 is satisfied.

It is known that the essential condition for a force acting between the fluid and the magnet\textbf{}ic field is the presence of an eddy current within the fluid, whereas the essential condition for the presence of eddy current is the presence of an eddy motional electromotive field \textbf{u}$\times$\textbf{B}. When both sides of Equation (1) are zero, the motional electromotive field within the fluid is irrotational. In this case, neither an eddy motional electromotive field nor an eddy current exists in the fluid.
Consequently, no magnetic force is exerted on the fluid when it crosses a magnetic field with velocity \textbf{u}.

Now, the circumstance that determines the conditions under which the fluid can freely traverse magnetic field without magnetic resistance turns into another circumstance that determines the conditions under which both sides of equation (1) equal zero.
To encounter the latter, we expand Equation (1) as

\begin{equation}
\frac{\partial{\textbf{B}}}{\partial{t}}=
\textbf{\textbf{u}}(\nabla\cdot\textbf{B})
-\textbf{\textbf{B}}(\nabla\cdot\textbf{u})
-(\textbf{\textbf{u}}\cdot\nabla)\textbf{B}
+(\textbf{\textbf{B}}\cdot\nabla)\textbf{u}.\label{e2}
\end{equation}

The first term on the right-hand side involves the divergence of the magnetic field $\nabla\cdot\textbf{B}$, which, according to Maxwell's equation, is always zero.

The factor $\nabla\cdot\textbf{u}$ in the second term is the divergence of the velocity field. If the density distribution of the fluid remains constant, this term also equals zero.

When each fluid element moves along a path of constant magnetic field \textbf{\textbf{B}} (namely, an isomagnetic surface), the third term $(\textbf{\textbf{u}}\cdot\nabla)\textbf{B}$ vanishes.

If the fluid elements that synchronously cross the same magnetic field line have the same velocity (i.e., do not exhibit differential movement), the fourth term $(\textbf{\textbf{B}}\cdot\nabla)\textbf{u}$ vanishes also. In general, the differential movement is related to a change in shape of the fluid element.

Equation (2) incorporates three essential factors that control the interaction between a fluid and a magnetic field:
\begin{enumerate}
\item The expansion or compression of the fluid, i.e., the change in density;
\item The differential movement of the fluid, i.e., the change in shape of the fluid; and
\item The change in magnetic field along the path of movement.
\end{enumerate}

These essential factors indicate that the so-called frozen-in phenomenon in reality relates to some changes. Without these changes, this phenomenon cannot develop.

If a fluid does not change its density distribution and shape, it can be regarded as a quasi-rigid body. The following simple law applies: \emph{When crossing a magnetic field along an isomagnetic surface, a rigid or quasi-rigid body will not be subject to any magnetic resistance, no matter how strong the magnetic field is and how fast the fluid moves.} This law fits the scenarios shown in Figs. 1, 2, and 3. The three scenarios share two characteristics:

\begin{enumerate}
\item All of the fluid elements cross the magnetic field along an isomagnetic surface.
\item The motional electromotive field \textbf{u}$\times$\textbf{B} inside the fluid is everywhere nonzero and curl-free.
\end{enumerate}

In the cases of Figs. 1 and 2, although the motional electromotive field \textbf{u}$\times$\textbf{B} inside the fluid is irrotational, a rotational  field \textbf{u}$\times$\textbf{B} exists at the boundary of the conductive body.
However, $\nabla\times(\textbf{u}\times\textbf{B})\neq0$ is only a necessary, but not sufficient, condition for the appearance of eddy currents.
The eddy plane at the boundary is always perpendicular to the boundary surfaces; furthermore, no conducting matter exists outside the boundary and eddy currents cannot cross the boundary surfaces.
This ensures that no eddy current can appear at the boundary and hence the magnetic field cannot be modified.

In the case of Fig. 3, the conductive body is a ring, and there has neither front boundary nor rear boundary in the direction of movement. The boundary analysis is appropriate only for the top and bottom surfaces of the conductive body.

For the case shown in Fig. 3, even if the outer and inner fluids rotate at different angular speeds, i.e. $\omega=f(r)$, no magnetic resistance acts on the fluid. However, if the differential rotation can be described as $\omega=f(z)$ ($z$ is the axial coordinate), a magnetic resistance will act on the fluid. This distinction arises from the fourth term of the right-hand side of Equation (2). If $\omega=f(r)$, then $(\textbf{\textbf{B}}\cdot\nabla)\textbf{u}=0$ and no magnetic force is felt; if $\omega=f(z)$, then $(\textbf{\textbf{B}}\cdot\nabla)\textbf{u}\neq0$ and a magnetic resistance is felt.
This result is consistent with the law of iso-rotation (Ferraro's theorem) which states that in the steady state, angular velocity is constant along magnetic lines.\citep{Alfven63} However, if $\omega=f(r)$, the angular velocity varies only along the direction perpendicular to the field lines.
In this situation, though the fluid is no longer quasi-rigid because of the appearance of differential rotation, it can still traverse magnetic field lines freely. This indicates that the constraint condition of ``quasi-rigid" can be softened to a certain extent.

Alfv\'{e}n \& F\"{a}lthammar has described an example similar to that in Fig. 1, for which the fluid can cross a uniform magnetic field under certain conditions\cite{Alfven63}. Our work extends this idea by softening the constraint conditions for other cases of non-uniform magnetic field and/or differential rotation of the fluids.

Because the case shown in Fig. 3 illustrates the situation in the vicinity of the equatorial plane of an aligned pulsar, we come to an important conclusion: the plasma near the equatorial plane of an aligned pulsar can cross the magnetic field and cannot be frozen with the magnetic field lines. In other words, the plasma can neither drive the magnetic field lines, nor be driven by the magnetic field lines. This conclusion is supported by the results of our magnetohydrodynamic experiments (see section ~\ref{section-experiments}).

\begin{figure}
\centerline{\includegraphics{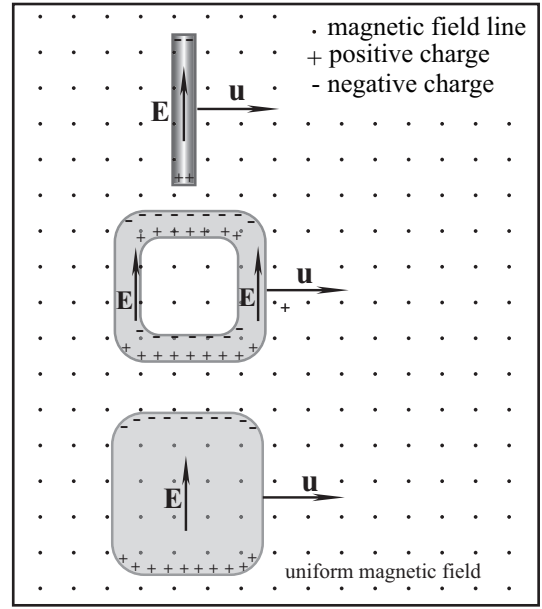}}
\caption{First non-frozen scenario. Moving across a uniform magnetic field, a fluid experiences no magnetic resistance. Here, the distributions of the charges and electric fields have been simplified.} \label{Fig1}
\end{figure}

\begin{figure}
\centerline{\includegraphics{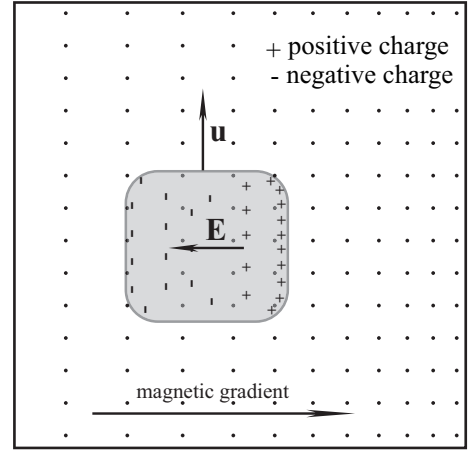}}
\caption{Second non-frozen scenario. For a fluid moving across a non-uniform magnetic field, there is no frozen-in effect provided the fluid elements move along the isomagnetic surface.} \label{Fig2}
\end{figure}

\begin{figure}
\centerline{\includegraphics{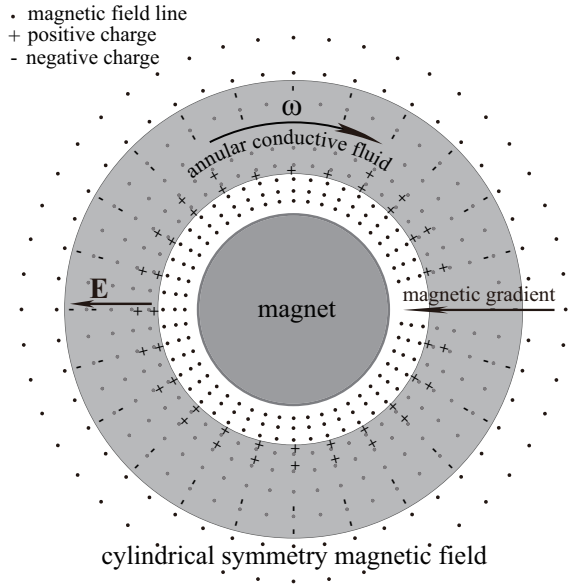}}
\caption{Third non-frozen scenario. In a cylindrically symmetric
magnetic field, the fluid rotation around the magnetic axis would not result in the frozen-in resistance, even if the outer and inner fluids rotate at different angular speeds. This scenario mirrors the situation near the equatorial plane of an aligned pulsar or the sun.} \label{Fig3}
\end{figure}

Figs. 1, 2, and 3 do not include all of the scenarios in which fluid can freely cross magnetic field lines. The conditions in these three scenarios merely ensure that all terms of the right-hand side of Eq. (2) are zero. More generally, even if some terms are non-zero, so long as the terms sum to zero, the fluid can freely cross the magnetic field lines. For example, if a fluid element undergoes an apropos expansion rate, it can cross magnetic field lines against the direction of magnetic gradient.

Fig. 4 shows a scenario in which the fluid experiences magnetic resistance and results in magnetic evolution. When combine Fig. 2 and Fig. 4 to Fig. 5, the significance of isomagnetic surface can be seen clearly.

If a magnetic field changes in the direction of fluid movement, the fluid will generally experience resistance. In particular, if the length of fluid along the direction of its motion is greater than the extent of the magnetic field, as shown in Fig. 6, magnetic resistance is inevitable at the boundary of the magnetic field because the magnetic field changes sharply there.

Now, it has been clear that when a quasi-rigid fluid crosses a uniform magnetic field, the system is always force-free, no matter what velocity the liquid has and how strong the magnetic field is. On the other hand, every steady magnetic field can be regarded as a superposition of uniform and non-uniform components and only the non-uniform component may, not must, result in magnetic resistance.
From this perspective, we can clearly understand the interaction between the fluid and the magnetic field and avoid to be misled by the frozen theorem.

\begin{figure}
\centerline{\includegraphics{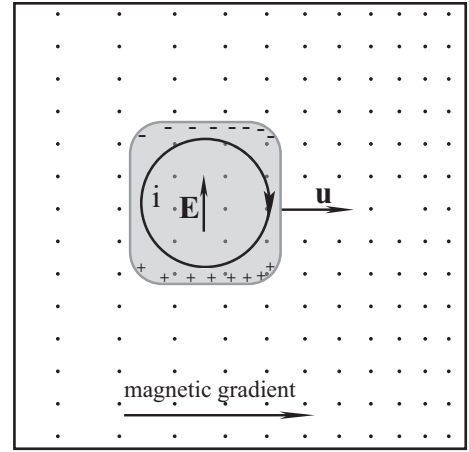}}
\caption{Magnetically frozen scenario. If the fluid velocity has a component in the direction of the magnetic field gradient, the fluid movement will result in a magnetic evolution and magnetic resistance. The field lines caused by the eddy current were not illustrated.} \label{Fig4}
\end{figure}
\begin{figure}
\centerline{\includegraphics{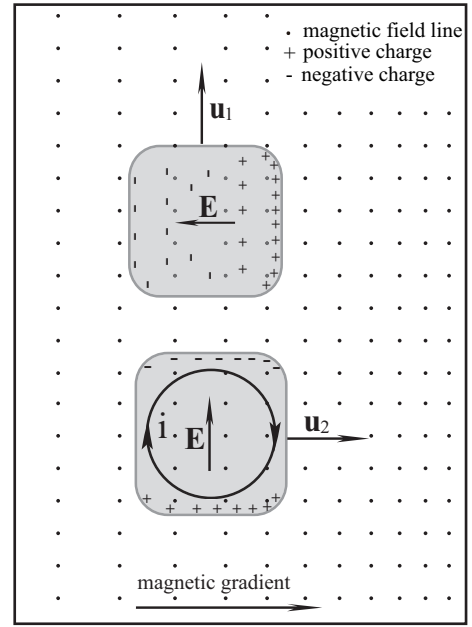}}
\caption{Significance of isomagnetic surface. The upper fluid can move freely because it moves along the isomagnetic surface. The lower fluid cannot do so. This figure is a composite of Figs. 2 and 4.} \label{Fig5}
\end{figure}
\begin{figure}
\centerline{\includegraphics{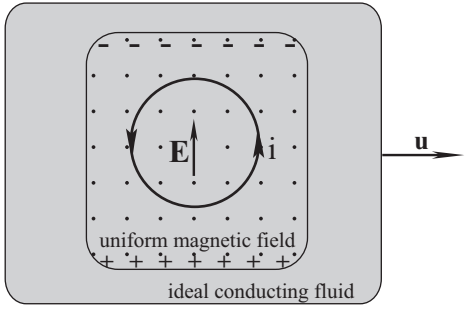}}
\caption{Another magnetically frozen scenario. If the extent of the magnetic field is smaller than that of the fluid, even if the magnetic field is uniform, frozen-in resistance will appear because of the high magnetic field gradient near the magnetic field boundary.} \label{Fig6}
\end{figure}

\subsection{Effects of the polarization charges and polarization electrostatic field}

The following equation describes the distribution of polarization charges:
\begin{equation}
\displaystyle{\frac{q}{\epsilon_0}}=-\nabla\cdot
(\textbf{u}\times\textbf{B})
\label{e5}
\end{equation}
In MHD literature, it seems that this equation is often belittled\citep{{GR95}} even discarded from the MHD equations.\cite{Bateman78}
This might arise from the concept that conducting fluid is always electrically neutral anywhere inside a conducting fluid.
In fact, the interaction between the fluid and the magnetic field can destroy the electrical neutrality and result in the appearance of surplus positive or negative charges inside the fluid. Especially, under the special conditions (such as Fig. 1, 2, and 3), the fluid's moving along the isomagnetic surface would result in neither magnetic disturbance nor magnetic force, but would result in only the electric field and charges to appear in the systems. In these special cases, only the effects of the electric field and charges are noteworthy. Their importance can be shown by two examples. First, in the magnetosphere of pulsars, the density distribution of charges has been studied by Goldreich \& Julian\citep{GJ69} and known by pulsar scientists as Goldreich-Julian density.\citep{Lyne05} Second,the output voltage of magnetohydrodynamic generator is just caused by the induced electric field.

This demonstrates that in some cases, the electrical neutrality approximation must be abandoned and the effects of the electric field and charges must be considered.

According to conventional physics, a straight wire, closed wire or conducting plate (see Fig. 1) moving perpendicularly in a uniform magnetic field can neither produce electric current nor experience resistance from the magnetic field. Thus, the magnetic field lines were not considered to be frozen within the conductors. Because this phenomenon is independent of the type or conductivity of the materials, even if the objects shown in Fig. 1 comprise a fluid such as mercury or plasma, the frozen-in effect does not apply. However, in the context of plasma physics, if the fluids in Fig. 1 are conductive, one can consider that fluids will be frozen in the magnetic field and cannot cross freely through the field lines. This presents an apparent contradiction between conventional physics and plasma physics.

In previous studies on the frozen-in effect, the effects of the polarization charges and electrostatic field \textbf{E} shown in Figs. 1, 2, and 3 have been disregarded.
Indeed, it is the electric field \textbf{\textbf{E}} that produces a fluid drift and ``thaws" the magnetic-frozen states. A simple calculation using the drift theory of charged particles\citep{Bittencourt04} verifies that the drift velocity of both the ions and electrons produced by \textbf{\textbf{E}} and \textbf{\textbf{B}} is always equal to the macroscopic moving velocity of the fluid, i.e., $\textbf{v}_{\rm{drift}}$=(\textbf{E}$\times$\textbf{B})/$\emph{B}^2\equiv$\textbf{u}.

Bellan\citep{Bellan06} has demonstrated that a plasma immersed in a static magnetic field \textbf{B} and a static electric field \textbf{E}, with \textbf{E} perpendicular to \textbf{B}, will freely drift across the magnetic field lines. Based on Bellan's viewpoint, we claim that the static electric field $\textbf{E}$, which is established by polarization charges, can naturally appear under certain conditions. The process by which this can occur is summarized below:

\begin{itemize}
  \item A fluid moving across the magnetic field lines induces an irrotational motional electromotive field \textbf{u}$\times$\textbf{B} inside the fluid;
  \item As a result of this irrotational motional electromotive field, polarization charges appear inside the fluid;
  \item The polarization charges result in a static polarization electric field \textbf{E}=$-\textbf{u}\times$\textbf{B};
  \item \textbf{E} combines with \textbf{B} to ``force" all the particles (ions and electrons) to drift with velocity $\textbf{v}_{\rm{drift}}$ equal to \textbf{u}. Namely, the velocity of the guiding-center drift equal to the motion velocity of the fluid.
\end{itemize}
Finally, the system is force-free and the fluid can move freely as the magnetic field is absent.

When an individual charged particle travels through a magnetic field, it is subject to Lorentz forces which compel it to gyrate around the magnetic field lines. However, within a fluid, the presence of both polarization charges and electric field create a very different situation. Under some special conditions, such as those illustrated in Figs. 1, 2, and 3, the static electric field produced by the polarization charges can counteract the Lorentz force and the fluid elements will drift across the magnetic field lines without any resistance.

Since fluid movement can result in two evolutions, when we cannot distinguish the crossing velocities by the magnetic evolution, we can still distinguish them by the distributions of charges or electric field.

\subsection{Infinite conductivity vs finite current}

There is an opinion in some textbooks:\citep{Bittencourt04, Li06, Ma12, Zheng09}
 \begin{quote}
 When an ideal conducting fluid traverses magnetic field lines, it must induce an electric field. However, because the conductivity is infinite and an infinite current is impossible, so the induced electric field and the perpendicular velocity component must be infinitesimally small and the field-line crossing motion is therefore prohibited.
\end{quote}
In fact, this opinion is wrong.

From electrical engineering we know that in a non-steady circuit system, beside resistance, inductive and capacitive reactance can also restrain the current growth. In other words, it is the complex impedance (not only the resistance) that controls the current growth. This law is also valid in MHD field because the electric field, current, and magnetic field are generally time-variant. It is similar that in the DC circuit, only resistance is considered, but for the AC circuit, the reactance also needs to be considered. Below, we show how the inductive reactance and capacitive reactance restrain the current growth and why the fluid can traverse magnetic field lines.

When a fluid traverses magnetic field lines, a motional electromotive field \textbf{u}$\times$\textbf{B} must appear to drive a current to grow. At the same time, an induced electric field\textbf{ E }immediately appears to restrain the current growth and makes the current always finite. The relation between them is $\textbf{E}=-\textbf{u}\times\textbf{B}$. Since the motional electromotive field \textbf{u}$\times$\textbf{B} has two components, rotational and irrotational components, the induced electric field \textbf{E} has two components too.

The rotational component of the motional electromotive field is $\nabla\times (\textbf{u}\times\textbf{B})$. The opposite component of the induced electric field is $\nabla\times \textbf{E}=-\partial \textbf{B}/\partial t$. This induced electric field component is caused by the time-variant magnetic field. The time-variant magnetic field is caused by the time-variant eddy current and this eddy current is caused by the rotational motional electromotive field. As long as the direction of $\nabla\times (\textbf{u}\times\textbf{B})$  does not reverse, the eddy current will grow continually. But the rate of growth depends on the inductive reactance. In a real system, the magnetic field caused by the eddy current only offsets the change of the external magnetic field along the path and conserves the magnetic flux; so the eddy current must be finite unless the external magnetic field can become infinitely strong. In addition, if a fluid moves along isomagnetic surface, the rotational component of the motional electromotive field must be equal to zero and the infinite eddy current is even more impossible.

The irrotational component of the motional electromotive field is $\nabla\cdot (\textbf{u}\times\textbf{B})$ and the opposite component of induced electric field is $\nabla\cdot \textbf{E}=q/\epsilon_0$. This induced electric field component is caused by the polarization charges and the polarization is driven by the irrotational motional electromotive field.
As long as a fluid traverses a magnetic field, this irrotational component of the induced electric field is always non-zero as shown in Fig. 1-5. In fact, the output voltage of magnetohydrodynamic generator is just caused by this irrotational component. This is the best evidence to prove that the induced electric field is generally non-zero.
Because relating usually to slow variation the capacitive reactance is very high, so that the capacitive current can usually be ignored. But the polarization electric field and charges are often important (see also the section ~\ref{section-discussion}).

It is correct to think always $\textbf{E}+(\textbf{u}\times \textbf{B})=0$, but it is wrong to think always $\textbf{E}=0$ and $\textbf{u}\times \textbf{B}=0$. When taking the inductive and capacitive reactances into account, we can understand why $\textbf{E}\neq 0$ and $\textbf{u}\times \textbf{B}\neq 0$ would not result in an infinite current inside ideal conducting fluid.

From the above analyses, we can get following logical relation: If a fluid cannot traverse magnetic field lines, the motional electromotive field $\textbf{u}\times \textbf{B}$  must be equal to zero. If $\textbf{u}\times \textbf{B}\equiv 0$, the induced electric field \textbf{E} must be equal to zero. If $\textbf{E}\equiv 0$, then $\partial \textbf{B}/\partial t\equiv 0$  and $q\equiv 0$. In this manner, the evolution equations $\partial \textbf{B}/\partial t=-\nabla\times(\textbf{u}\times\textbf{B})$ and $q=\epsilon_0\nabla \cdot \textbf(\textbf{u}\times\textbf{B})$  will be insignificant.

All in all, the non-zero motional electromotive field caused by the fluid's traversing motion is the root of the system evolution and would not result in an infinite current. Supposing that the fluid freezes with the field lines and deny the field-line crossing, we will actually deny the system evolution. Therefore we have to accept the concept that a conducting fluid can traverse magnetic field lines. The field-line crossing is not only acceptable but also indispensable.

\subsection{Problems arising from the frozen theorem}

\begin{figure}
\centerline{\includegraphics{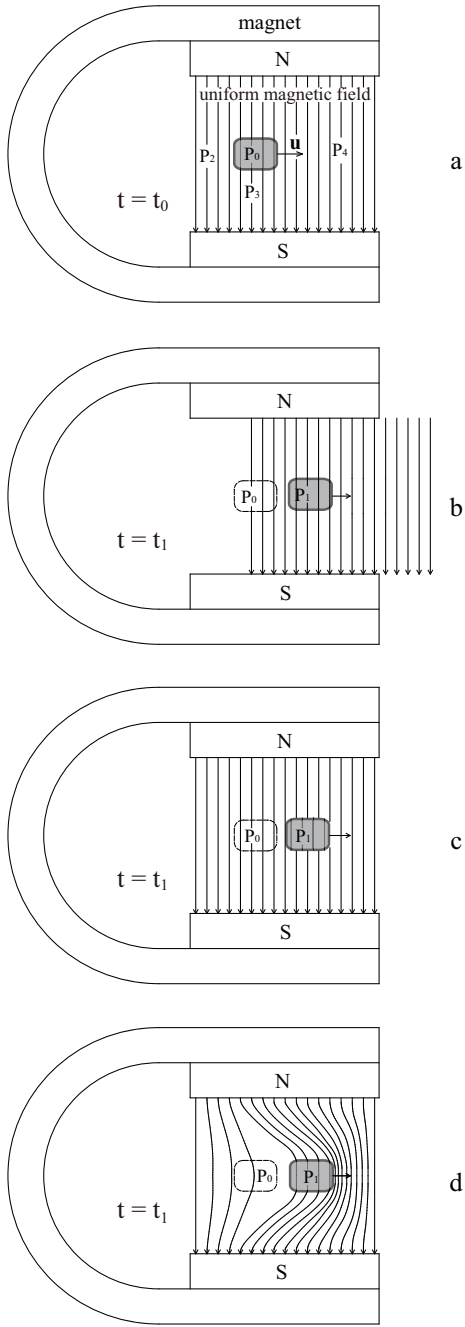}}
\caption{Different scenarios for changing magnetic field lines outside a fluid. (a) The fluid will move across the uniform magnetic field. (b) If the fluid can carry the field lines inside and outside the fluid with the same velocity, the field lines will translate and retain their shape. (c) If the fluid can carry only the field lines inside the fluid, the field lines will be repeatedly cut off and reconnected at the fluid boundary. (d) If the fluid can carry the field lines inside the fluid and all field lines remain continuous, the field will become non-uniform.} \label{fig7}
\end{figure}

\begin{figure}[b]
\centerline{\includegraphics{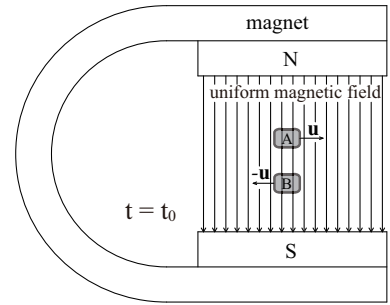}}
\caption{Fluids A and B are originally fixed on common field lines. When they move in opposite directions, how will the common magnetic field lines change?} \label{Fig8}
\end{figure}

We have discussed above in regard to Fig. 1 that, given a uniform magnetic field and a quasi-rigid fluid, the fluid elements can freely cross magnetic field lines. Other researchers,\citep{Stern66} however, have demonstrated that if the fluid elements are fixed on the magnetic field lines and move always together, Equation (1) is equally well satisfied.
Some researchers have thus advocated that both choices are valid: \citep{Northrop63} 1) the fluid moves across magnetic field lines; and
2) the fluid and the magnetic field lines move together (i.e. frozen theorem). However, following careful analysis, some problems will arise from the frozen theorem.

In previous discussions of the frozen theorem, attention has usually been devoted to only the fluid and the magnetic field lines inside the fluid. However, as long as the system studied is extended to also include the magnet and the magnetic field lines outside the fluid (see Fig. 7), we will see that contradictions are unavoidable.

First, we assume that the magnetic field is uniform, the magnet is immobile and the quasi-rigid fluid moves with velocity \textbf{u} relative to the magnet, as shown in Fig. 7a. Then we consider the changes of the magnetic field lines outside the fluid.

If the field lines outside the fluid respond to the fluid motion, only three possible solutions can be considered, as illustrated in Fig. 7b, 7c, and 7d.

\begin{enumerate}
\item In Fig. 7b, all magnetic field lines outside the fluid translate with the fluid. As the velocities are identical everywhere, the configuration of the field lines cannot change and the field lines must be cut off at the boundary of the magnet. Apparently, this solution is unacceptable.
\item In Fig. 7c, all the magnetic field lines outside the fluid are stationary relative to the magnet. In this scenario, the field lines must be cut off by the fluid boundary and must repeatedly reconnect. However, this result contradicts the alternative representation of the frozen theorem: ``Any fluid element that is at one instant on a magnetic field line will be on the original magnetic field line at any other instant". Because a reconnected magnetic field line is not the original one, then the fluid cannot remain attached to the original magnetic field lines.
    If we accept that the magnetic field lines can be cut off, we have really changed the frozen theorem to that \emph{any field-line crossing of the fluid will cut off the field lines}.
    Apparently, this solution is also unacceptable.
\item In Fig. 7d, the magnetic field lines outside the fluid are dragged by the fluid and change the magnetic field configuration. However, this configuration induces a magnetic evolution and resistance, inconsistent with the analysis of subsection A. Therefore, it, too, should be abandoned.
\end{enumerate}

Another scenario is illustrated in Fig. 8. Fluids A and B are originally fixed on common field lines. When the two fluids move in opposite directions, changes of the field lines, especially the common ones, cannot be dealt with successfully with the help of the frozen theorem.

With the help of the analyses above, we conclude that, as long as the system studied is extended and the magnet and its field lines outside the fluid are also included, the frozen theorem always leads to contradictions.

\subsection{Better interpretation than the frozen theorem}

\begin{figure}
\centerline{\includegraphics{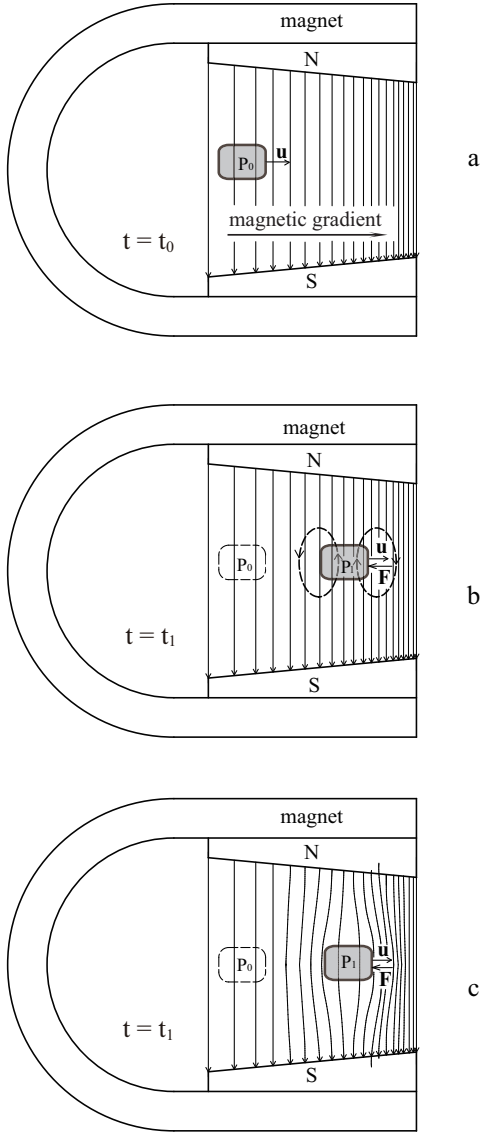}}
\caption{Two-magnet interpretation  for the interaction between fluid and magnetic field lines. When moving from P$_0$ to P$_1$, the fluid becomes a magnet, as shown in panel b. When the two magnetic fields in panel b are combined, the magnetic configuration is as illustrated in panel c.} \label{Fig9}
\end{figure}

\begin{figure}
\centerline{\includegraphics{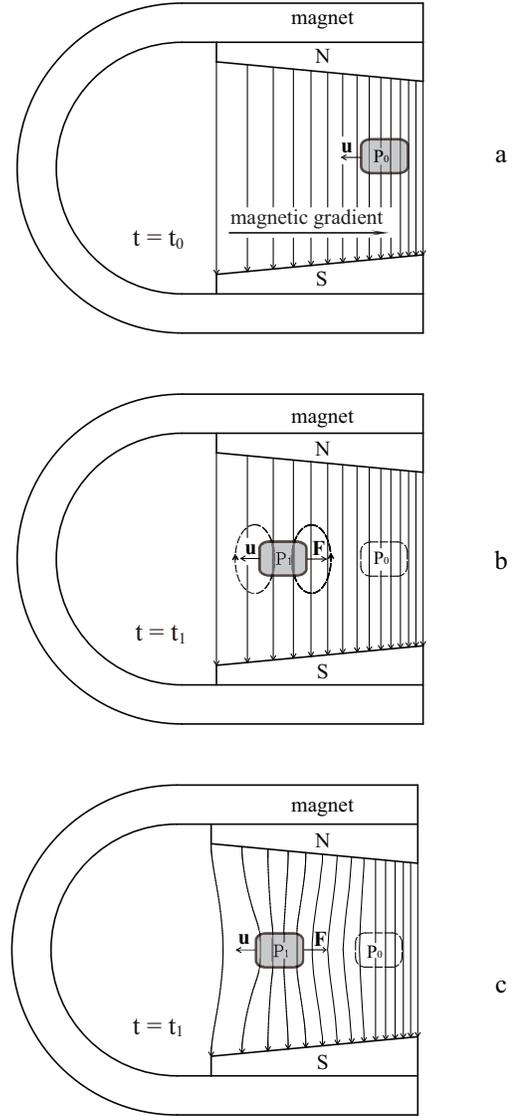}}
\caption{The opposite of Fig. 9. The fluid moves backward with the magnetic gradient.} \label{Fig10}
\end{figure}

We suggest using a better method to describe the relationship between moving fluid and magnetic field lines. When a fluid crosses a magnetic field along its isomagnetic surface, we can consider that the field lines are stationary with the magnet and the fluid crosses them freely.
If a fluid crosses a non-uniform magnetic field along the direction of magnetic gradient, as shown in Fig. 9a, we can still consider that the field lines caused by the magnet are stationary, but the magnetic gradient creates an eddy current inside the fluid and the fluid has become a fluidic magnet. The field lines associated with this fluidic magnet are shown with dashed lines in Fig. 9b.

The magnetic resistance is, in fact, the force acting between the external magnet and the fluidic magnet. Although the forces acting on the front and rear sides of the fluid are always opposite, the magnetic gradient makes their magnitudes different. No matter what direction the fluid is moving, parallel or opposite to the direction of magnetic gradient (respectively in Figs. 9 and 10), the result of the forces is always opposite to the motion direction.

When the two magnetic fields combine to form one field, Fig. 9b becomes Fig. 9c. The magnetic flux in the fluid remains constant but the magnetic configuration outside the fluid has changed compared with Fig. 9a.

A notable point is that the dashed field lines at P$_1$ in Fig. 9b are produced in P$_1$ instead of pulled from P$_0$, because the magnetic field lines had not appeared when the fluid was at P$_0$. Therefore, this two-magnet interpretation needs neither the concept of frozen-in field lines nor the motion of field lines.

Under the condition that the fluid is a quasi-rigid body, this two-magnet interpretation infers the following:
\begin{enumerate}
\item{When a fluid crosses magnetic field lines, eddy current can appear inside the fluid.}
\item{The eddy current turns the fluid into a fluidic magnet and causes an additional magnetic field to appear.}
\item{Because the two-magnets interact with each other, the magnetic resistance appears.}
\item{The additional magnetic field always counteracts changes in magnetic flux caused by the external magnet and conserves the total flux inside the fluid.}
\item{The additional magnetic field will change the magnetic configuration outside the fluid.}
\item{Because the eddy current depends on the magnetic gradient, the effects above disappear outright if the fluid moves along the isomagnetic surface.}
\end{enumerate}

Compared with the frozen theorem, this two-magnet interpretation has the following strengths:

\begin{enumerate}

\item{Using the two-magnet interpretation, we can clearly understand not only the magnetic conservation inside the fluid but also the magnetic variation outside the fluid.}
\item{ The concept of magnetic reconnection has no longer been needed, even on the fluid boundary.}
\item{The problems shown in Figs. 7b--d can be avoided altogether.}
\item{Since this two-magnet interpretation emphasizes that the eddy current depends on the magnetic gradient, the importance of the isomagnetic surface become easier to understand.}
\item{Since this two-magnet interpretation emphasizes the field-lines crossing, it comes clear that the fluid motion causes the volume density distribution of the polarization charges.}
\end{enumerate}

Note that this interpretation is very different with the frozen theorem. As shown in Fig. 9c, when the fluid moves from P$_0$ to P$_1$, the field lines to the left of P$_1$ grow denser instead of sparser. If the frozen theorem holds and field lines have been carried by the fluid from P$_0$ to P$_1$, the field lines there must become sparser and are similar to Fig. 7d.
Similarly, as shown in Fig. 10c, when the fluid moves left to location P$_1$, ahead of the fluid, the field lines do not become denser; quite the reverse, they become sparser.
This indicates that this two-magnet interpretation is not compatible with the frozen theorem. In other words, from the two concepts, we cannot derive identical evolution results for the magnetic field lines outside the fluid.
Therefore, it is impossible maintain the two concepts concurrently and one must be abandoned.

\subsection{Summary contrasts in table}
\begin{table*}[t]
\caption{\label{tab:tableI}Contrast between two cases of Eq. (1)}
\begin{ruledtabular}
\begin{tabular}{lll}
&case A&case B\\ \hline
both sides of Equation (1)&= 0&$\neq 0$ \\\hline
property of magnetic field&steady magnetic field&time-varying magnetic field\\\hline
eddy current in fluid&absence&presence\\\hline
magnetic resistance&absence&presence\\\hline
illustration&Figs. 1, 2, and 3&Figs. 4 and 6\\
\end{tabular}
\end{ruledtabular}
\end{table*}

\begin{table*}
\caption{\label{tab:tableII}Summary of the analysis results of Eqs. (1) and (3)}
\begin{ruledtabular}
\begin{tabular}{lll}
&magnetic freezing (column A)&magnetic ``thaw''(column B)\\ \hline
correlative equations&$ \partial{\textbf{B}}/\partial{t}=\nabla\times
(\textbf{u}\times\textbf{B})$&$q/\epsilon_0=-\nabla\cdot
(\textbf{u}\times\textbf{B})$ \\\hline
motional electromotive field parameter&curl&divergence\\\hline
induction phenomena&inductive eddy current&polarization charges and electric field \\\hline
correlative magnetic field parameter&magnetic field gradient&magnetic field strength \\\hline
direction of fluid movement&parallel to magnetic gradient&perpendicular to magnetic gradient\\\hline
effect of magnetic force&resisting the traversing motion&inducing field-line traversing\\\hline
illustration&Figs. 4 and 6&Figs. 1, 2, and 3\\
\end{tabular}
\end{ruledtabular}
\end{table*}

\begin{table*}[!htb]
\caption{\label{tab:tableIII}Summary of the expressions of the frozen theorem}
\begin{ruledtabular}
\begin{tabular}{p{7cm}p{7cm}}
correct expressions&incorrect expressions\\ \hline
The magnetic flux in any fluid element is conserved.
&Any motion of fluid, perpendicular to the field lines, carries the field lines with the fluid.\\ \hline
Any two elements of the same fluid that are at one instant on a common magnetic field line will be on a common magnetic field line at any other instant.\citep{Falthammar06, GR95}&
Any fluid element that is at one instant on a magnetic field line will be on the original magnetic field line at any other instant .\\\hline
 Quasi-rigid fluid can freely traverse magnetic field lines only along the isomagnetic surface.
&A fluid cannot freely traverse magnetic field lines.
\end{tabular}
\end{ruledtabular}
\end{table*}

Table~\ref{tab:tableI} contrasts two cases and clarifies the relationship between the electromagnetic phenomena and Eq. (1).

For clarity, the analysis results of Eqs. (1) and (3) are summarized in Table~\ref{tab:tableII}.
Previous studies have mainly focused on those properties listed under column A and have overlooked all of the aspects listed in column B; that is, the frozen-in effect has been emphasized whereas the drift effect has been largely ignored.
Consequently, the frozen theorem was often stated in the form \emph{a fluid cannot freely cross magnetic field lines.}

The frozen theorem was expressed in several forms, which are summarized in Table~\ref{tab:tableIII}. All of the expressions listed in the right-hand column are derived from the frozen theorem, and hence are incorrect or incomplete.

\section{\label{section-experiments}Experiments on the isomagnetic surface}

\begin{figure}
\centerline{\includegraphics{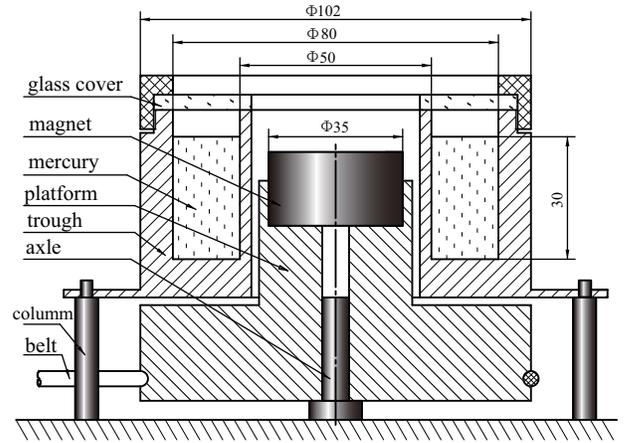}}
\caption{Schematic figure of drive experimental device. The unit of length is millimeter. The belt drives the rotation of the platform and magnet; the other components are stationary. } \label{Fig11}
\end{figure}

To verify the special significance of isomagnetic surface, we conducted a series of magnetohydrodynamic experiments. The configuration is the same as that of Fig. 3. For simplicity and clarity, the experiments were recorded and the videos of these can be viewed on YouTube. Figs. 11 and 12 depict the device configurations and the sizes of the main components.
Apart from the magnet, all components are made of non-magnetic material (plastic or copper). The mercury was placed in a cylindrical trough surrounding a rotating platform. At the magnet surface, the magnetic field intensity is about 0.6 T. The rotational rates are shown in the relevant videos.

The experiments were divided into two groups.
The first group comprised three pairs of experiments to determine the conditions under which the magnet can drive rotation of the mercury. In each experiment the basis of comparison is an \emph{aligned magnetic rotator}, a cylindrically symmetric magnet whose magnetic axis aligns with its rotation axis. The three experiments are as follows:

\begin{enumerate}
\item \emph{Drive experiment A}\cite{footnote1} revealed the contrast between the aligned rotator and a rotator with magnetic axis orthogonal to the rotation axis. This contrast test indicates that the transverse magnetic field can drive the mercury to rotate, but the aligned magnetic field cannot do that.

\item \emph{Drive experiment B}\cite{footnote2} revealed the contrast between the aligned rotator and a magnet with magnetic axis parallel to, but displaced from the rotation axis. This contrast test indicates that the eccentricity magnetic field can drive the mercury, but the coaxial magnetic field cannot do that.

\item \emph{Drive experiment C}\cite{footnote3} revealed the contrast between the aligned rotator and a magnet that is not cylindrically symmetric. This contrast test indicates that the cylindrically asymmetric magnetic field can drive the mercury, but the cylindrically symmetric magnetic field cannot do that.
\end{enumerate}

\begin{figure}
\centerline{\includegraphics{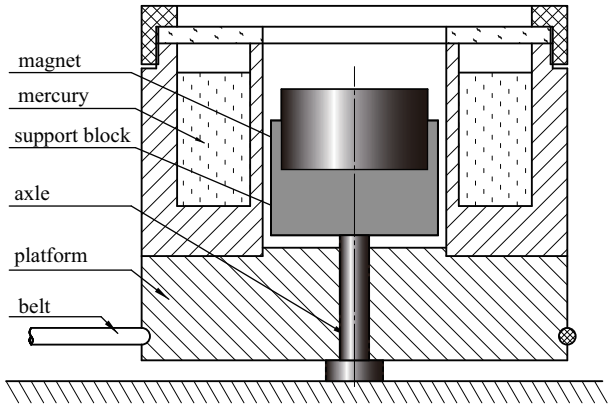}}
\caption{Schematic figure of braking experimental device. The axis, support block, and the magnet are stationary. The belt drives the rotation of other parts.} \label{Fig12}
\end{figure}

The first group of experiments demonstrated that an aligned magnetic rotator cannot drive the mercury to rotate. However, this result has not proved the absence of the magnetic frozen effect in such cases, since the same result also arise if the magnetic field lines do not co-rotate with the aligned magnet\cite{footnote7}. The value of these experiments is to answer the question: Can an aligned pulsar drive its plasma and achieve the co-rotation state? The answer is NO. This result denied clearly the co-rotating model of aligned pulsar.

In the second group of experiments, the conditions
under which rotating mercury can be slowed by a stationary magnet were determined.
The mercury trough (rather than the central magnet) was spun until the
mercury revolution reached a steady state. In each experiment the
basis of comparison was an \emph{aligned magnet}, a cylindrically
symmetric magnet whose magnetic axis matched that of the mercury
rotation axis. In this case, the mercury eventually
revolved at the same rate as the trough. For the three comparison cases,
braking forces reduced the rotation of the mercury. We used the
same magnets in every pair of braking
experiments, designated A, B and C and described below.

\begin{table*}[!htb]
\caption{\label{tab:tableIV}Summary of the experimental results}
\begin{ruledtabular}
\begin{tabular}{lcc}
&Driving experiment results&Braking experiment results\\ \hline
aligned axis and cylindrical symmetry cases&no\footnotemark[1]&no\\\hline
orthogonal magnetic field &yes\footnotemark[2]&yes\\\hline
off-axis magnetic field&yes&yes \\\hline
cylindrically asymmetric magnetic field&yes&yes
\end{tabular}
\end{ruledtabular}
\footnotetext[1]{ ``no'' implies the absence of magnetic resistance. }
\footnotetext[2]{ ``yes'' implies the presence of magnetic resistance. }
\end{table*}

\begin{enumerate}
\item \emph{Braking experiment A}\cite{footnote4} revealed the contrast between
the aligned magnet and the orthogonal magnet.
This contrast test indicates that the orthogonal magnetic field can slow down the mercury, but the aligned magnetic field cannot do that.

\item \emph{Braking experiment B}\cite{footnote5} revealed the contrast between
the aligned magnet and the off-axis magnet.
This contrast test indicates that the eccentricity magnetic field can slow down the mercury, but the aligned magnetic field cannot do that.

\item \emph{Braking experiment C}\cite{footnote6} revealed the contrast between
the aligned magnet and the cylindrically asymmetric magnet.
This contrast test indicates that the cylindrically asymmetric magnetic field can slow down the mercury, but the cylindrically symmetric magnetic field cannot do that.

\end{enumerate}

The braking experiments demonstrated that even if the magnetic field is non-uniform, fluid can experience no frozen-in effect, as long as all of the fluid elements move along isomagnetic surface. Thereby, the analyses for Fig. 3 have been corroborated by our experimental results.

Table~\ref{tab:tableIV} presents a summary of the experimental results.
These experimental results have verified the special significance of isomagnetic surface: Along isomagnetic surface, a magnetic field can neither drive nor brake conducting fluid.

Some may think that our experimental results are obvious and worthless, However, we believe, at least, that the most pulsar scientists have not observed and understood such experimental results. Otherwise, they would not believe that a pulsar can drive its upper plasma to co-rotate, even if the magnetic and spin axes are aligned.\citep{GJ69}

\section{\label{section-discussion}Discussion}

Our ideas are embodied in the following quote from Feynman et al.\citep{Feynman64}: ``It makes no sense to say something like: When I move a magnet, it takes its field with it, so the lines of \textbf{B} are also moved. There is no way to make sense, in general, out of the idea of `the speed of a moving field line.'"

A magnetic field essentially cannot be ascribed a movement.
Time-varying magnetic fields exist, whereas moving ones do not. While it can be intuitive and useful to regard magnetic field lines as moving, this approach may lead to erroneous results.

In conventional physics, certain electromagnetic problems have been correctly solved by assuming moving field lines. But in those situations, the magnetic poles always move with the field lines. By contrast, typical discourses on the frozen theorem do not clarify whether the magnetic source body is in motion or at rest. If the external magnet is at rest (generally the default condition), the assumption that the field lines are movable always makes the physics more complex and even leads to wrong results. Herein lies the root of the problem caused by the frozen theorem.

For some astrophysical objects, the magnetic field can be very
strong, but the magnetic gradient in the direction of the plasma movement is usually very small, especially for cases similar to that shown in Fig. 3. Consequently, the effect of magnetic resistance should be
much weaker than that expected and the plasma can in fact freely cross the magnetic field lines. In other words, in some astrophysical objects, the magnetic frozen-in effect can be ignored provided that the magnetic gradient along the path of plasma movement can be ignored.

Magnetic confinement technology has been widely applied in nuclear fusion engineering. In those devices, plasma is restrained in the magnetic beam and cannot move laterally. One interpretation for this phenomenon is that the very strong magnetic field restrains the plasma; actually, the transverse magnetic gradient does the restraining, especially near the boundary surface of the magnetic beam. Any uniform magnetic field, regardless of strength, cannot restrain a conductive substance.
But some people think that at low beta where the magnetic field is strong, the fluid will be pushed around by the magnetic field lines and if the field is stationary, so is the fluid. The reason why some people think so may be that they were misled by the frozen theorem.

Formerly, the plasma was considered usually as an electrically neutral matter. The analysis above indicates that the interaction between plasma and magnetic field can destroy the electrical neutrality. If the existence of the polarization charges is considered, some problems can become very simple.
For example, the origin of the magnetic field of sunspots can be simply interpreted using our new theory.

Here, we suppose that the solar general magnetic field does not rotate with the sun\cite{footnote7}
and accept that plasma can cross the magnetic field lines along the isomagnetic surface. Fig. 3 then can illustrate the charge distribution near the sun's equatorial plane.
It can therefore be understood that the plasma of the photosphere is no longer electrically neutral.
When the surplus positive or negative charges are carried by the plasma and revolve around the center of the plasma typhoon, the eddy electrical current and the magnetic fields must appear and accompany the sunspots.
We have checked the relationships among the direction of the sun's rotation, the direction of the sun's general magnetic field, the electrical polarity of photosphere, the direction of the plasma typhoon's rotation, and the direction of the magnetic field of the sunspots. We are sure that these relationships are consistent with the electromagnetic theory.

The magnetic field of sunspots can be simply explained as follows: 1) It is the sun's rotation that results in plasma crossing the non-rotating general magnetic field along the isomagnetic surface; 2) It is the field-line crossing motion that results in polarization charges appearing in the photosphere; 3) It is the eddy current of the polarization charges that results in the magnetic field of the sunspots. From these relationships, we can clearly see the important significance of polarization charges and isomagnetic surface.

In the earth's atmosphere, there is vertical electric field and hence there are polarization charges. Therefore, typhoon's magnetic field should be detected as in sunspots. It may be a method to test our theory.
\section{summary}
\begin{enumerate}
\item When the magnetic evolution and magnetic force are the objects of study, the electrical neutrality is indeed an excellent approximation. However, when the drift of fluid particles and fluid's crossing motion are concerned, we must pay more attention to the electric field and the charges distribution, and in this case the electrical neutrality approximation is unacceptable.
\item We emphasize the important significance of the isomagnetic surface and polarization charges to understand entirely and accurately the interaction between conducting fluid and magnetic field.
\item If we accept the existence of polarization charges and the polarization electric field, we must accept the existence of the \textbf{E}$\times$\textbf{B} drift and the fact that the fluid can generally cross magnetic field lines; hence, we must reconsider the frozen theorem.
\item Whether a quasi-rigid conducting fluid can freely traverse magnetic field lines or not depends solely on the magnetic gradient along the fluid path. No matter how strong the magnetic field is or how fast the fluid speed is, a quasi-rigid fluid can freely traverse the magnetic field lines as long as it is along the isomagnetic surface. Especially, no matter how strong the magnetic field is, a uniform magnetic field cannot push around the fluid.
\item The literal meaning of ``frozen-in" is liable to promote an erroneous association between fluid traversing motion and magnetic resistance. Therefore, we suggest that the use of this term should be avoided whenever possible. In particular, the frozen theorem should not be indiscriminately used to study dynamics problems.

\end{enumerate}


\begin{thebibliography}{aip}

\bibitem[Akasofu(1978)]{Akasofu78} S.-I. Akasofu,
\textit{Space Sci. Rev.} \textbf{21}, 489 (1978).

\bibitem[Alfv\'{e}n(1942)]{Alfven42} H. Alfv\'{e}n,
\textit{Nature} \textbf{150}, 405 (1942).

\bibitem[Alfv\'{e}n(1958)]{Alfven58}  H. Alfv\'{e}n,
\textit{Tellus}  \textbf{10}, 104 (1958).

\bibitem[Alfv\'{e}n \& F\"{a}lthammar(1963)]{Alfven63} H. Alfv\'{e}n and C.-C. F\"{a}lthammar,
\textit{Cosmical electrodynamics, fundamental principles,} 2nd ed. (Clarendon Press, 1963), p. 110, p.190 Fig. 5.10.

\bibitem[Alfv\'{e}n(1976)]{Alfven76} H. Alfv\'{e}n,
\textit{J. Geophys. Res.} \textbf{81}, 4019 (1976).
\bibitem[Bateman(1978)]{Bateman78} G. Bateman,
\textit{MHD Instabilities}, (Cambridge University Press 1978), p. 30.


\bibitem[Bellan(2006)]{Bellan06} P. M. Bellan,
\textit{Fundamentals of plasma physics,} (Cambridge University Press 2006), p. 100.

\bibitem[Bittencourt(2004)]{Bittencourt04} J. A. Bittencourt,
\textit{Fundamentals of plasma physics,} 3rd ed. (Springer 2004), p. 312, p. 51, p. 315.

\bibitem[Boyd \& Sanderson(2003)]{Boyd03} T. J. M. Boyd and  J. J. Sanderson,
\textit{The physics of plasmas.} (Cambridge University Press 2003), p. 79.

\bibitem[F\"{a}lthammar(2006)]{Falthammar06} C.-G. F\"{a}lthammar,
\textit{Am. J. Phys.} \textbf{74}, 454 (2006).

\bibitem[Feynman et al.(1964)]{Feynman64} R. P. Feynman, R. B. Leighton, and M. Sands,
\textit{The Feynman Lectures on Physics}, Vol. 2 (Addison wesley 1964), p. 13-10.

\bibitem[Goldreich \& Julian(1969)]{GJ69} P. Goldreich and W. H. Julian,
\textit{Astrophys. J.} \textbf{157}, 869 (1969).





\bibitem[Goldston \& Rutherford(1995)]{GR95} R. J. Goldston and P. H. Rutherford,
\textit{Introduction to plasma physics}, (Institute of Physics Publishing 1995), P. 118, P.124.






\bibitem[Ivonin et al.(1998)]{Ivonin98} I. A. Ivonin, V. P. Pavlenko, and H. Persson,
\textit{Phys. Plasmas} \textbf{5}, 2893 (1998).

\bibitem[Li et al.(2006)]{Li06} D. Li, Y. H. Chen, J. X. Ma, and W. H. Yang,
\textit{Plasma Physics,} (Higher Education Press 2006), p. 69, Chinese.

\bibitem[Lyne \& Graham-Smith(2005)]{Lyne05} A. Lyne and F. Graham-Smith,
\textit{Pulsar Astronomy,} 3nd ed. (Cambridge University Press, 2005), p. 23.

\bibitem[Liang \& Liang(2009)]{footnote1} Z. X. Liang, and Y. Liang, http://www.youtube.com/watch?v=\\YCc4ybfKiY0, YouTube (2009).

\bibitem[Liang \& Liang(2009)]{footnote2} Z. X. Liang, and Y. Liang, http://www.youtube.com/watch?v=\\ZiVNxVAqUKc, YouTube (2009).

\bibitem[Liang \& Liang(2009)]{footnote3} Z. X. Liang, and Y. Liang, http://www.youtube.com/watch?v=\\SBAzUBzc2xM, YouTube (2009).

\bibitem[Liang \& Liang(2009)]{footnote4} Z. X. Liang, and Y. Liang, http://www.youtube.com/watch?v=\\C7DxRoVdtxw, YouTube (2009).

\bibitem[Liang \& Liang(2009)]{footnote5} Z. X. Liang, and Y. Liang, http://www.youtube.com/watch?v=\\p9UzfT6UHcE, YouTube (2009).

\bibitem[Liang \& Liang(2009)]{footnote6} Z. X. Liang, and Y. Liang, http://www.youtube.com/watch?v=\\Eq0B5XR\_o8U, YouTube (2009).

\bibitem[Lundin et al.(2005)]{Lundin05} R. Lundin, M. Yamauchi, J.-A. Sauvaud, and A. Balogh,
\textit{Annales Geophysicae} \textbf{23}, 2565 (2005).

\bibitem[Ma et al.(2012)]{Ma12} J. C. Ma, X. W. Hu, and Y. H. Chen,
\textit{Fundamentals of plasma physics,} 2nd ed. (University of Science and Technology of China Press 2012), p. 188, Chinese.




\bibitem[Northrop (1963)]{Northrop63}T. G. Northrop, \textit{Rev. Geophys}, \textbf{1}, 283 (1963).

\bibitem[Stern (1966)]{Stern66}D. P. Stern, \textit{Space Science Reviews}, \textbf{6}, 147 (1966).

\bibitem[Zheng(2009)]{Zheng09} C. K. Zheng,
\textit{Plasma Physics,} (Peking University Press 2009), p. 79, Chinese.

\bibitem{footnote7}When a magnet rotates around its own magnetic axis, whether the magnetic field corotates with the magnet has been a controversial issue. This issue is also named as Faraday paradox. Although most of astrophysicists think the magnetic field corotates with the magnet, we take the opposite point of view because it is accepted by the most of general physical scientists and supported by more experiments for Faraday paradox.





\end{thebibliography}
\end{document}